\documentclass{article}

\PassOptionsToPackage{numbers, compress}{natbib}

   \usepackage[preprint]{neurips_2020}

\usepackage[utf8]{inputenc} 
\usepackage[T1]{fontenc}    
\usepackage{hyperref}       
\usepackage{url}            
\usepackage{booktabs}       
\usepackage{amsfonts}       
\usepackage{nicefrac}       
\usepackage{microtype}      
\usepackage{bm} 
\usepackage{amsmath,amssymb}
\usepackage{graphicx}
\graphicspath{{./figures/}}
\usepackage{subcaption}
\usepackage{enumitem}
\usepackage{tikz}
\usepackage{soul} 

\newlength{\imwidth}
\newtheorem{theorem}{Theorem}

\renewcommand{\vec}[1]{\bm{#1}}
\DeclareMathOperator*{\argmin}{arg\,min}
\DeclareMathOperator{\sign}{sign}

\newcommand{\Spm}{\vec{S}_{\pm}}
\newcommand{\Szero}{\vec{S}_{0}}

\newcommand{\x}{{\vec{x}}}
\newcommand{\y}{{\vec{y}}}
\newcommand{\z}{{\vec{z}}}

\newcommand{\e}{{\vec{e}}}
\newcommand{\params}{{\vec{\theta}}}
\newcommand{\W}{{\vec{W}_\params}}
\newcommand{\Wno}{{\vec{W}}}
\newcommand{\Q}{{\mathcal{Q}}}
\newcommand{\J}{{\mathcal{J}}}
\newcommand{\R}{\mathcal{R}}

\newcommand*\diff{\mathop{}\!\mathrm{d}}

\title{Supervised Learning of Sparsity-Promoting Regularizers for Denoising}

\author{%
  Michael T.~McCann\\
  Department of Computational Mathematics, Science and Engineering\\
  Michigan State University\\
  East Lansing, MI 48824 \\
  \texttt{mccann13@msu.edu} \\
  \And
  Saiprasad Ravishankar\\
  Department of Computational Mathematics, Science and Engineering\\
  Michigan State University\\
  East Lansing, MI 48824 \\
  \texttt{ravisha3@msu.edu}
}

\begin{document}

\maketitle

\begin{abstract} 
We present a method for supervised learning
of sparsity-promoting regularizers for image denoising.
Sparsity-promoting regularization
is a key ingredient in solving modern image reconstruction problems;
however, the operators underlying these regularizers  are usually either
designed by hand
or
learned from data in an unsupervised way.
The recent success of supervised learning 
(mainly convolutional neural networks) in solving image reconstruction problems
suggests that it could be a fruitful approach to designing regularizers.
As a first experiment in this direction,
we propose to denoise images using a variational formulation with a parametric, sparsity-promoting regularizer, 
where the parameters of the regularizer are learned to minimize the mean squared error of reconstructions on a training set of (ground truth image, measurement) pairs.
Training involves solving a challenging bilievel optimization problem;
we derive an expression for the gradient of the training loss using Karush–Kuhn–Tucker conditions and
provide an accompanying gradient descent algorithm to minimize it.
Our experiments on a simple synthetic, denoising problem
show that the proposed method can learn an operator that
outperforms well-known regularizers
(total variation, DCT-sparsity, and unsupervised dictionary learning)
and collaborative filtering.
While the approach we present is specific to denoising,
we believe that it can be adapted to the whole class of inverse problems with linear measurement models,
giving it applicability to a wide range of image reconstruction problems.
\end{abstract}

\section{Introduction}
Image reconstruction problems, where an image must be recovered from its noisy measurements,
appear in a wide range of fields, including 
computer vision,
biomedical imaging,
astronomy,
nondestructive testing,
remote sensing,
and geophysical imaging;
see \cite{bertero_introduction_1998} for a textbook-length introduction.
or \cite{mccann_biomedical_2019,ravishankar_image_2020} for tutorial-length introductions.
One prominent approach to solving these problems is to formulate reconstruction as an optimization problem---%
seeking the image that best fits the measured data according to a prescribed model.
We refer to this approach as variational reconstruction because it relies on
the calculus of variations to find extrema.
These models typically comprise a model of the imaging system
(including noise) and a model of plausible images.
We restrict the current discussion to objective functions of the form
$\mathcal{J}(\x) = \| \vec{H} \x - \y \|_2^2 + \R(\x)$,
where $\x \in \mathbb{R}^n$ is a (vectorized) image,
$\y \in \mathbb{R}^m$ are its noisy measurements, 
$\vec{H} \in \mathbb{R}^{m \times n}$ is a linear forward model,
and $\R: \mathbb{R}^n \to \mathbb{R}$
is a regularization functional.
While this formulation implicitly assumes
a linear imaging system and Gaussian noise,
it is nonetheless applicable to many modern image problems.

A prominent theme in designing the regularization functional, $\R$,
has been that of sparsity:
the idea that the reconstructed image $\x$ should admit a sparse 
(having a small number of nonzero elements)
representation in some domain.
Examples of these models would be
synthesis sparsity ($\x \approx \Wno \z$ and $\z$ is sparse)~\cite{tropp_greed_2004},
analysis sparsity ($\Wno \x$ is sparse)~\cite{rubinstein_analysis_2013},
and transform sparsity ($\Wno\x \approx  \z$ and $\z$ is sparse)~\cite{ravishankar_learning_2013}.
A common and successful approach to promoting sparsity
in image reconstructions
is to use a regularization functional of the form $\|\Wno \x\|_1$,
where $\Wno \in \mathbb{R}^{k \times n}$
is called a sparsifying transform,
analysis dictionary,
or analysis operator---%
we adopt the latter terminology.
The resulting objective functions are usually convex,
and
results from compressive sensing \cite{candes_robust_2006}
show that, under certain conditions,
the $\ell^1$ penalty
provides solutions with exactly sparsity.
While there are several choices for $\Wno$
that work well in practice
(e, g. wavelets, finite differences, Fourier or discrete cosine transform (DCT)),
several authors have also sought to learn $\Wno$ from data,
an approach called dictionary (or transform) learning ~\cite{tosic_dictionary_2011,rubinstein_analysis_2013,ravishankar_learning_2013}.

While variational approaches have dominated the field of image reconstruction for the past decades,
a recent trend has been to train supervised methods,
especially convolutional neural networks (CNNs),
to solve image reconstruction problems.
Recent reviews~\cite{mccann_convolutional_2017,ongie_deep_2020} give a good picture of this evolving field,
but we want to point out a few trends here.
Many recent papers attempt to incorporate some aspects of variational reconstruction into a supervised scheme.
One way of doing that is by
learning some aspect of the regularization,
e.g., via learned shrinkage operators \cite{helor_discriminative_2008,shtok_learned_2013,kamilov_learning_2016,nguyen_learning_2018},
learned plug-and-play denoisers~\cite{alshabili_learning_2020},
using a CNN to enforce a constraint-type regularization ~\cite{gupta_cnn_2018},
or allowing a CNN to take the part of the regularizer in an unrolled iterative  algorithm~\cite{aggarwal_modl_2019}.
Broadly speaking,
these are attempts to combine the benefits of variational reconstruction
(theoretical guarantees, parsimony)
with the benefits of supervised learning (data adaptivity, state-of-the-art performance).
The approach in this work falls on the same spectrum,
but is decidedly on the ``shallow learning'' end---%
we aim to bring the benefits of supervision to variational reconstruction,
rather than the other way around.

We propose to solve image reconstruction problems using a sparsifying analysis operator that is learned from training data in a supervised way.
Our learning problem is 
\begin{subequations}
\begin{equation}
    \argmin_\params \Q(\params), \text{\quad where \quad}
    \Q(\params) = \sum_{t=1}^T \frac{1}{2} \|\x^*(\W, \y_t) - \x_t \|_2^2,
    \label{eq:main_upper}
\end{equation}
where
$\W \in \mathbb{R}^{k\times n}$ is a sparsifying operator parameterized by $\params \in \mathbb{R}^p$,
$\x_t$ and $\y_t \in \mathbb{R}^n$ are the $t$th training image and its corresponding noise-corrupted version,
  \begin{equation}  
   \quad \x^*(\Wno, \y) = \argmin_\x 
    \frac{1}{2}\| \x - \y \|_2^2 + \beta \| \Wno \x \|_1
    \label{eq:main_lower},
\end{equation}%
\label{eq:main}%
\end{subequations}%
and 
$\beta \in \mathbb{R}$ is a fixed, scalar parameter
that controls the regularization strength during reconstruction.
Note that while \eqref{eq:main_upper} refers to a specific parameterization of the sparsifying operator, $\W$,
\eqref{eq:main_lower} is defined for any operator $\Wno$,
regardless of parameterization;
we will continue to use $\W$ when the role of the parameterization is important and $\Wno$ otherwise.
Equation \eqref{eq:main_lower} is a variational formulation of image reconstruction for denoising (i.e., $\vec{H} = \vec{I}$),
where we have made the dependence on the operator $\Wno$ explicit.
Problem \eqref{eq:main_upper} is the minimum mean squared error (MMSE) minimization problem on the training data.
In statistical terms,
the interpretation of \eqref{eq:main} is that we are seeking a prior (parameterized by $\params$)
so that the MAP estimator (given by \eqref{eq:main_lower})
approximates the MMSE estimator~\cite{gribonval_reconciling_2013}.
Because we aim to work on images, we assume that $\W$ is too large to fit into memory,
but rather is implemented via functions for applying $\W$ and its transpose to vectors of the appropriate size.
This is true, e.g., when $\W$ implements convolution(s),
as is the case in our experiments.

\paragraph{Related Work}
A 2011 work \cite{peyre_learning_2011} poses
the problem \eqref{eq:main} except with a differentiable
version of $\|\cdot\|_1$
and provides experiments for one-dimensional signals.
In \cite{mairal_task_2012}, the authors address a synthesis version of \eqref{eq:main},
wherein the reconstruction problem involves finding sparse codes, $\z$ such that $\|\x - \Wno\z \|$ is small.
This change from the analysis to synthesis formulation means that the optimization techniques used in \cite{mairal_task_2012} do not apply here.
In \cite{sprechmann_supervised_2013}, the authors derive gradients for a generalization of \eqref{eq:main} by relaxing $\|\Wno\x\|_1$ to $\alpha\|\Wno\x - \z \|_2^2 + \|\z\|_1$.
This approach gives the gradient in the limit of $\alpha \to \infty$,
however the expression requires computing the eigendecomposition of a large matrix.
Therefore the authors use the relaxed version $\alpha < \infty$ in practice.
In a brief preprint~\cite{chen_learning_2014},
the authors derive a gradient for \eqref{eq:main}
by expressing $\Wno$ in a fixed basis and 
using a differentiable relaxation.
Finally, \cite{chen_insights_2014}
provides a nice overview of the topic analysis operator learning in its various forms,
and also tackles \eqref{eq:main} using a differentiable sparsity penalty.

\paragraph{Contribution}
The main contribution of this work is to derive a gradient scheme for minimizing \eqref{eq:main} without relying on relaxation.
Smooth relaxations of the penalty term can result in slow convergence%
~\cite{peyre_learning_2011},
which our formulation avoids.
Our formulation has the added advantage of promoting exact sparsity%
~\cite{candes_robust_2006}
and also has a close connection to CNNs:
the regularization functional in \eqref{eq:main_lower} can be expressed as
a shallow CNN with rectified linear unit (ReLU)
activation functions.
This connection indicates that we may be able to generalize the methods here to learn a more complex CNN-based regularizer.
The computational bottleneck of our approach is solving the reconstruction problem \eqref{eq:main_lower} at each step of gradient descent on \eqref{eq:main_upper};
we propose a warm-start scheme to ameliorate this cost,
allowing training to finish in hours on a consumer-level GPU.
In a denoising experiment,
we demonstrate that the sparsifying operator learned using our method
provides improved performance over popular fixed operators,
as well as an operator learned in an unsupervised manner.

\section{Methods}
In order to apply gradient-based techniques to solving \eqref{eq:main},
our first goal is to compute the gradient of $\Q$
with respect to $\params$.
The challenge in this lies in computing
the partial derivatives (with respect to $\params$) of the elements of each vector $\x^*_t$
(i.e., the Jacobian matrix of $\x^*_t$),
after which we can obtain the desired result using the chain rule.
Again we note that we expect $\W$ to be too large to fit into memory,
thus our derivation does not rely on factoring $\W$.
Our approach is first to develop 
a closed-form expression 
(i.e., not including $\argmin$) for $\x^*(\Wno, \y)$,
and then
to derive the desired gradient.

\subsection[Closed-form Expression for x*(W, y)]
{Closed-form Expression for $\x^*(\Wno, \y)$}

Consider the functional 
\begin{equation}
     \label{eq:analysis}
       \J(\x, \Wno, \y) = \frac{1}{2}\| \x - \y \|_2^2 + \beta \| \Wno \x \|_1.
 \end{equation}
It is strictly convex in $\x$
(because the $\ell^2$ norm term is strictly convex
and the $\ell^1$ norm term is convex)
and therefore has a unique global minimizer.
Thus we are justified in writing $\x^*(\Wno, \y) =\argmin_\x \J(\x, \Wno, \y)$
without the possibility of
the minimizer not existing or being a nonsingleton.
Note that although $\x^*$ depends on $\y$, $\beta$ and $\Wno$,
the $\y$- and $\beta$-dependence is not relevant for this derivation and
we will not continue to notate it explicitly.

Our key insight is that we can have a closed-form
expression for $\x^*$ in terms of $\Wno$
if we know the sign pattern of $\Wno\x^*(\Wno)$,
and we can always find the sign pattern of $\Wno\x^*(\Wno)$ by solving \eqref{eq:main_lower}.
We first define some notation.
Let $\vec{c}(\Wno)$ denote the sign pattern
associated with a given $\Wno$,
i.e.,
$\vec{c}(\Wno) = \sign(\Wno \x^*(\Wno))$,
where $[\sign(\z)]_i$ 
is defined to be -1 when $[\z]_i < 0$;
0 when $[\z]_i = 0$;
and 1 when $[\z]_i > 0$.
Considering a fixed $\vec{c_0} = \vec{c}(\Wno_0)$,
we define matrices that pull out the rows
of $\Wno$ that give rise to zero, negative, and positive values in $\Wno \x^*(\Wno)$.
Let $k_{=0}$, $k_{\ne0}$, $k_{<0}$, and $k_{>0}$ denote the number of
zero, nonzero, negative, and positive elements of $\vec{c}_0$, respectively.
Similarly, let $[\vec{\pi}_0]_m$, $[\vec{\pi}_{<0}]_m$, and $[\vec{\pi}_{>0}]_m$
denote the index of the $m$th zero, negative, and positive element of $\vec{c}_0$.
Let $\Szero \in \mathbb{R}^{k_{=0} \times k}$ and $\Spm \in \mathbb{R}^{k_{\ne0} \times k}$
be defined as
\begin{equation}
    [\Szero]_{m,n}=  \begin{cases}
      1 &\quad \text{if} \quad [\pi_0]_m = n;\\
      0 &\quad \text{otherwise};
\end{cases}
\text{\quad and \quad}
    [\Spm]_{m,n}=  \begin{cases}
      1 &\quad \text{if} \quad [\pi_{>0}]_m = n;\\
      -1 &\quad \text{if} \quad [\pi_{<0}]_{m-k_{>0}} = n;\\
      0 &\quad \text{otherwise.}
\end{cases}
\end{equation}

With this notation in place, we can write that
\begin{equation}
  \label{eq:linear-transform-constrained}
  \x^*(\Wno) = 
  \argmin_{\vec{x}} \frac{1}{2}\| \vec{x} - \vec{y} \|_2^2 + 
  \beta \|\Wno \vec{x} \|_1, \quad
   \text{s.t.} \quad \Szero\Wno\x = \vec{0}
\end{equation}
for all $\Wno$ such that $\vec{c}(\Wno) = \vec{c}_0$.
This is true because 
whenever $\vec{c}(\Wno) = \vec{c}_0$,
the minimizer of \eqref{eq:analysis}
is feasible for \eqref{eq:linear-transform-constrained}.
Similarly,
we use $\Spm$ to simplify the $\ell^1$ norm term,
\begin{equation}
  \label{eq:linear-transform-local}
  \x^*(\Wno) = 
  \argmin_{\vec{x}} \frac{1}{2}\| \vec{x} - \vec{y} \|_2^2 
  + \beta \vec{1}^T \Spm \Wno \vec{x}, \quad
   \text{s.t.} \quad \Szero\Wno\x = \vec{0},
\end{equation}
where $\vec{1}$ is a $k_{\ne0}\times 1$ vector of ones,
and, again the equality holds for all $\Wno$ such that $\vec{c}(\Wno) = \vec{c}_0$.

Now that we have transformed the problem into an equality-constrained quadratic minimization,
we can use standard results  
(e.g., see \cite{boyd_convex_2004} Section 10.1.1)
to state the KKT conditions for \eqref{eq:linear-transform-local}:
\begin{equation}
  \label{eq:linear-transform-KKT}
  \underbrace{
  \begin{bmatrix}
    \vec{I} &  \Wno^T \Szero^T \\
    \Szero\Wno & \vec{0}
  \end{bmatrix}}_{\vec{A} \in \mathbb{R}^{(n + k_{=0}) \times (n + k_{=0})}}
\underbrace{
  \begin{bmatrix}
    \x^*(\Wno) \\
    \vec{\nu}
  \end{bmatrix}}_{\vec{z} \in \mathbb{R}^{n + k_{=0}}}
=
\underbrace{
  \begin{bmatrix}
    \y - (\beta \vec{1}^T \Spm \Wno)^T \\
    \vec{0}
  \end{bmatrix}}_{\vec{b} \in \mathbb{R}^{n+k_{=0}}},
\end{equation}
where the underbraces give names ($\vec{A}$, $\vec{z}$, and $\vec{b}$) to each quantity
to simplify the subsequent notation.
Because $\vec{I}$ is nonsingular,
$\vec{A}$ is invertible whenever $\Szero\Wno$ has full row rank~\cite{boyd_convex_2004}.

In order to cleanly write the result,
we define two more selection matrices that are useful for
pulling out the $\x$ and $\vec{\nu}$ parts of $\vec{z}$.
Let $\vec{P}^\x \in \mathbb{R}^{n \times (n + k_{=0})}$ and
$\vec{P}_{\vec{\nu}}  \in \mathbb{R}^{k_{=0} \times (n + k_{=0})}$
be defined as
$\vec{P}_\x =
  \begin{bmatrix}
  \vec{I} & \vec{0} \end{bmatrix}$
and
$
\vec{P}_{\vec{\nu}} =
  \begin{bmatrix}
   \vec{0} & \vec{I}
\end{bmatrix}.
$
We can then state the following result:
\begin{theorem}[Solutions of sparse analysis-form denoising]
\label{thrm:solutions}
  For any $\vec{c}_0 \in \{-1, 0, 1\}^k$,
  and for all $\Wno$ such that $\sign(\Wno \x^*(\Wno)) = \vec{c}_0$
  with $\Szero\Wno$ full row rank, 
  the minimizer of the sparse analysis-form denoising problem \eqref{eq:analysis}
  is given by
    $\x^*(\Wno) = \vec{P}_\x \vec{A}^{-1} \vec{b}$,
    where each term in the right hand side depends on $\Wno$ as described above.
\end{theorem}
Theorem~\ref{thrm:solutions} provides a closed-form expression for $\x^*(\Wno)$ 
that is valid in each region where $\vec{c}(\Wno)$ is constant.
Because our formulation results in exact sparsity,
these regions can have nonempty interiors,
allowing us to compute gradients.
As a brief example of this, consider the scalar denoising problem
$x^*(w) = \argmin_x \frac{1}{2}(x-y)^2 + |w x|$.
Assuming that $y\ge0$,
one can show that $x^*(w) = y-|w|$ when $y-|w| \ge 0$ and $0$ otherwise.
As a result, $c((0, y)) = 1$, $c((-y, 0)) = -1$, and $c((-\infty, -y] \cup 0 \cup [y,\infty))=0$;
a similar result holds when $y\le0$.
Thus $\Q(w)$ is smooth except at $w=0,-y,y$, which form a set of measure 0.

Also, note that Theorem~\ref{thrm:solutions} could be extended to cover the case where 
$\vec{A}$ is not full rank by expressing $\x^*$
in terms of the pseudoinverse of $\vec{A}$.
In that case, the following derivations would need to be adapted 
to use the gradient of the pseudoinverse~\cite{golub_differentiation_1973}.


\subsection[Gradient of Q(θ) With Respect to θ]{Gradient of $\Q(\params)$ With Respect to $\params$}
Even using automatic differentiation software tools
(e.g., PyTorch~\cite{paszke_pytorch_2019}),
Theorem~\ref{thrm:solutions} does not provide a way to compute the gradient of $\Q$
in \eqref{eq:main} with respect to $\params$
because of the presence of $\vec{A}^{-1}$.
Due to the size of $\vec{A}$ in practice,
the inverse needs to be computed iteratively,
and tracking gradients though hundreds of iterations would require an impractical amount of memory.
To avoid this problem, 
we use matrix calculus to manipulate the gradients into a form
amenable to automatic differentiation.
In the following, we use the notation of \cite{minka_old_2000}
(a resource we highly recommend for this type of math),
where $\diff \y$ is defined to be the part of $\y(\x + \diff \x) - \y(\x)$
that is linear in $\diff \x$.
As in \eqref{eq:main} we use the subscript $t$ to denote
quantities that depend on the $t$the training pair.
From Theorem~\ref{thrm:solutions} and matrix calculus rules,
we have
\begin{equation}
\label{eq:grad_step_1}
    \diff\x^*_t = \vec{P}_{\x, t} 
    (-\vec{A}_t^{-1} \diff\vec{A}_t 
    \underbrace{\vec{A}_t^{-1}\vec{b}_t}_{\z_t} 
    +
    \vec{A}_t^{-1}\diff\vec{b}_t)
    \quad\textrm{and}\quad
    \diff \Q = \sum_{t=1}^T \diff{\x_t^*}^\intercal
    (\underbrace{\x_t^* - \x_t}_{\e_t}).
\end{equation}
Therefore,
\begin{equation}
    \diff \Q = \sum_{t=1}^T 
    (-\z_t^\intercal\diff\vec{A}_t 
    +
    \diff\vec{b}_t^\intercal)\vec{A}_t^{-1} \vec{P}_{\x,t}^\intercal \e_t,
    \label{eq:gradient}
\end{equation}
with $\vec{z}_t$ and $\vec{e}_t$ as defined in \eqref{eq:grad_step_1}
and where all gradients are with respect to $\params$.
As we will describe in the next section, 
it turns out that, with the help of automatic differentiation software, \eqref{eq:gradient} is sufficient to compute the desired gradient.

\subsection{Implementation of Gradient Calculation}
We now give an outline of how we compute this gradient in practice.
As we have stated before, 
we expect $\W$ to be too large to fit in memory.
As a result, our desired gradient is not with respect to $\W$,
but actually with respect to $\params$;
luckily, automatic differentiation software can handle this detail for us.
For a given $\vec{\theta}$ and for each $t$:
\begin{enumerate}[wide]
    \item Solve the reconstruction problem \eqref{eq:main_lower} to find $\x_t^*$.
    This can be accomplished via established techniques for convex optimization;
    we use the ADMM~\cite{boyd_distributed_2011} with the split variable $\z=\W \x$.
   
    \item Determine the selection matrices $\Szero$ and $\Spm$.
    Because the $\x_t^*$ obtained via ADMM may still have some small error,
    there may not be exact zeros in $\W\x_t^*$,
    which complicates determining $\Szero$.
    Our approach is to look for zeros in the ADMM split variable corresponding to 
    $\W\x_t$ because it is both approximately equal to $\W\x_t^*$ and because it has exact zeros
    (because it is the result of soft thresholding in ADMM).
    Alternatively,
    setting some small threshold on the magnitude $\W\x_t^*$ would have a similar effect.
 
    \item Solve the KKT system \eqref{eq:linear-transform-KKT} to find the $\z_t$ in \eqref{eq:grad_step_1}. 
    We use the conjugate gradient (CG) algorithm~\cite{shewchuk_introduction_1994}.
    While $\vec{A}$ is symmetric and therefore CG could be applied to solve $\vec{A}\z = \vec{b}$ directly, 
    we find that, because there could be some inexactness in $\x_t^*$,
    it is more stable to solve $\vec{A}^\intercal\vec{A}\z = \vec{A}^\intercal\vec{b}$.
    
    \item Compute the $\e_t$ in \eqref{eq:grad_step_1} and find $\vec{q}_t = \vec{A}^{-1}_t \vec{P}_{\x,t}^\intercal \e_t$
    by solving a linear system 
    $\vec{A}_t^\intercal\vec{A}_t\vec{q}_t = \vec{A}_t^\intercal \vec{P}_{\x,t}^\intercal \e_t$.
    
    \item Turn on automatic differentiation with respect to $\vec{\theta}$,
    compute the scalar quantity
     $   \vec{Q}_t = 
    (-\z_t^\intercal \vec{A}_t
    +
    \vec{b}^\intercal_t)\vec{q}_t$,
    and perform an automatic gradient calculation.
    The key here is that there are several quantities in \eqref{eq:gradient}
    that depend on $\vec{\theta}$ (through application of $\W$),
    but we only want automatic differentiation to happen through 
    $\vec{A}_t$ and $\vec{b}_t$, as indicated by \eqref{eq:gradient}.
\end{enumerate}
The total gradient is then the sum of the gradients for each $t$.
(In fact, rather than looping over $t$,
the whole process can be done in a single shot
by concatenating the $\y_t$'s into a long vector
when solving the reconstruction problem in Step 1;
this may be faster, but requires careful bookkeeping in the code.)

In summary, to compute the gradient at a given $\vec{\theta}$ (equivalently, $\W$),
we need to run ADMM once and CG twice.
The most expensive operation is the application of $\W$,
which happens repeatedly during CG and ADMM.
In our experiments, we needed a few hundred applications of $\W$
during CG  and a few thousand in ADMM, 
making the ADMM the bottleneck.

\subsection[Learning W]{Learning $\W$}
\label{sec:learning}
With gradient in hand,
we can use any of a variety of first-order methods to solve \eqref{eq:main},
including gradient descent,  stochastic gradient descent (SGD), ADAM~\cite{kingma_adam_2014}, or BFGS~\cite{avriel_nonlinear_2003}.
We found that SGD worked well in our experiments.

We make two additional implementation notes.
First, it is very important to monitor the accuracy of the inverse problems being solved
during gradient evaluation,
i.e., ADMM and CG,
and to ensure that their hyperparameters are such that accurate solutions are computed.
If the results are inaccurate,
gradients will be inaccurate and,
in our experience, learning will fail.
Second,
we found it useful to store the $\x_t^*$'s at each iteration
and use them as a warm start in the next iteration.
This allows fewer iterations to be used when solving the reconstruction problem,
thereby speeding up training.
(In the case of using ADMM to solve the reconstruction problem,
we need to save not only the $\x_t^*$'s,
but also the associated split and dual variables.)

\section{Experiments}
We now present the details of our proof-of-concept denoising experiments.
\subsection{Data and Evaluation} 
Our experiment uses simple, synthetic images
(see Figure~\ref{fig:denoising_results_gt} for an example).
The images are $64\times64$~pixels and generated according to a dead leaves model
that produces images that mimic the statistics of natural images~\cite{lee_occlusion_2001}.
Specifically,
an image is formed by superimposing 
a fixed number of rectangles (100, in our case),
each with a random height, width, and intensity (uniform between 0 and 1).
For each image, we create a noisy version by adding
IID, zero-mean Gaussian noise
with a standard deviation of 0.1
(see Figure~\ref{fig:denoising_results_noisy} for an example).
We use small (compared to typical digital photographs)
images to reduce training times;
we use the dead leaves model because it contains structure that we expect our approach will be able to capture.

Our experiments compare the ability of several methods (Section~\ref{sec:meth_compared})
to perform denoising---%
to restore a clean image from its noisy version.
For evaluation, we generate a testing set of ten images,
along with their corresponding noisy versions.
Our figure of merit is the SNR
on the entire testing set, defined as
\begin{equation}
\label{eq:SNR}
    10 \log_{10}\left(
    \sum_{t=1}^T \sum_{i=1}^n [\x_t]_i^2
    \right)
    - 10 \log_{10}\left( 
    \sum_{t=1}^T \sum_{i=1}^n ([\hat{\x}_t]_i  - [\x_t]_i)^2
    \right),
\end{equation} 
where $\hat{\x}_t$ is reconstruction of the $t$th noisy image in the testing set,
$\x_t$ is the $t$th ground truth image in the testing set,
and $[\x]_i$ is the $i$th pixel of $\x$.
The SNR is expressed in decibels (dB).
Note that the SNR as defined here is a monotone function of 
the training objective \eqref{eq:main_upper},
so the supervised method maximizes SNR on the training set.
Another common metric is average SNR
(i.e., SNR 
computed separately for each image
 and averaged),
which does not have this property.
This detail is unlikely to matter much, especially when the testing set is as homogeneous as ours.
In a setting where it does matter, $\eqref{eq:main}$ could be adapted to account for it.
Another common metric is the peak SNR (PSNR).
If one considers 1.0 to be the peak,
SNR values can be converted to PSNR values
by adding 
$10 \log_{10}\left(
    \sum_{t=1}^T \sum_{i=1}^n 1
    \right) - 
10 \log_{10}\left(
    \sum_{t=1}^T \sum_{i=1}^n [\x_t]_i^2
    \right)$;
this value is 4.69 dB on our testing set.


\subsection{Methods Compared}
\label{sec:meth_compared}
We compare the denoising performance of five methods.
Here, we briefly describe each method
including its parameters;
we discuss parameter tuning in the next section.

\textbf{BM3D}~\cite{makinen_exact_2019} denoises images patches
by combining similar patches into groups and performing sparse representation on the groups.
Its main parameter is the standard deviation of the noise,
with a higher value giving smoother results.
BM3D typically shows excellent performance on Gaussian denoising of natural images,
and we expect it to be strong on our dataset, as well.
We use the Python implementation from the author's website,
\url{http://www.cs.tut.fi/~foi/GCF-BM3D/}.

\textbf{Total variation (TV)} is a variational method that denoises 
by solving the the reconstruction problem \eqref{eq:main_lower} with
$\Wno$ fixed to a finite differencing operator.
The dimensions of $\Wno$ are $2(63)^2 \times 64^2$ because there are two filters
(vertical and horizontal differences) and the size of the valid convolution along each dimension is 64 - 2 + 1.
We use the anisotropic version of TV (with no 2-norm on the finite differences)
because it is a good fit for dataset and
to make it more comparable to the other methods.
We solve the reconstruction problem using the
ADMM~\cite{boyd_distributed_2011},
with 400 outer and 40 inner iterations,
which is sufficient for the cost to be
stable to 4 significant digits.
The TV method has one scalar parameter, $\beta$,
which controls the regularization strength.
Because TV promotes piecewise-constant reconstructions,
we expect it to perform well in the experiments on our dataset.

\textbf{DCT sparsity}, like TV,
is a variational method with a fixed $\Wno$.
In this case, $\Wno$ performs a 2D DCT on each $3\times3$ block of the input
(see Figure~\ref{fig:response_DCT} for the corresponding filters).
Following recent work with this type of regularizer~\cite{rubinstein_analysis_2013},
we remove the constant filter from the DCT,
meaning that $\Wno$ is $8(62)^2 \times 64^2$.
Like TV, the DCT method has one scalar parameter, $\beta$.


\textbf{Unsupervised learned analysis sparsity}
learns $\W$ by minimizing $\sum_{t=1}^T \|\W\x_t\|_1$
with respect to $\params$.
The learned $\W$ is then used in $\eqref{eq:main}$ to perform reconstruction.
We use the term \emph{unsupervised} because the $\y_t$'s are not used
during training.
We parameterize $\W$ by its filters,
so $\W$ is $9(62)^2 \times 64^2$
and $\params$ is $9(9) \times 1$;
these filters are initialized to the $3\times3$ DCT.
To avoid the trivial solution $\W=\vec{0}$, 
we constrain the filters of $\W$ to be
orthogonal during learning%
~\cite{yaghoobi_noise_2012,ravishankar_sparsifying_2015}.
We minimize the training objective using the
ADMM~\cite{boyd_distributed_2011}
with split variables $\z_t = \W\x_t$ and
where orthogonality is enforced by solving 
an orthogonal Procrustes problem using 
a singular value decomposition (SVD) each iteration~\cite{ravishankar_sparsifying_2015}.
After learning, 
we remove the first (constant) filter.
The main parameter of the algorithm is the $\beta$
used during reconstruction.
We expect that this method should be able to learn
a good sparsifying analysis operator because our dataset is very structured.

\textbf{Supervised learned analysis sparsity (proposed)} denoises by solving the reconstruction problem \eqref{eq:main_lower}
after $\W$ has been learned by supervised training,
i.e., solving \eqref{eq:main} on a training set.
We parameterize $\W$ as a convolution with eight,
learnable $3\times3$ filters,
initialized to the (nonconstant) $3\times3$ DCT filters.
The parameters of the supervised method are $\beta$
and the training hyperparamters.
If $\W$ were a global minimizer of \eqref{eq:main},
and if we assume the model generalizes from training to testing,
then the supervised method would perform optimal
 denoising of the testing set (in the MSE sense).
Unfortunately, \eqref{eq:main} is a challenging,
nonconvex problem, 
where we are likely to find only local optima in practice.

\subsection{Training and Parameter Tuning}
For training and parameter tuning,
we generate a dataset of ten training images
along with their corresponding noisy versions.
For the methods with only a scalar parameter
(BM3D, TV, and DCT),
we perform a parameter sweep and select the value
that maximizes the signal-to-noise ratio (SNR) of the results
on this training set.
For the unsupervised method,
we use the $\beta$ found for the DCT method and train $\W$
to sparsify clean images (as described in the previous section).
After training $\W$,
we perform another 1D parameter sweep on $\beta$,
finding the optimal value for denoising the training set.

For the supervised method,
we use the $\beta$ found during the DCT parameter sweep,
and perform stochastic gradient descent on $\eqref{eq:main}$,
with the gradient normalized by the number of pixels and batch size
and with with step size set to 2.0.
Following \cite{smith_dont_2017},
we increase the batch size during training to improve convergence.
Specifically, we use a batch size of 1 for 5000 iterations,
5 for 2500 more iterations,
and 10 for 2500 more iterations.
We do not adapt $\beta$ after training as
(unlike in the unsupervised case)
$\beta$ appears in the training objective
and therefore $\W$ should adapt to it.
Following best practices in reproducibility~\cite{sinha_designing_2020},
we note that we did perform multiple testing evaluations during the development of the algorithm.
These were mainly for debugging and exploring different initialization and learning schedules.
We did not omit any results that would change our reported findings.

\section{Results and Discussion}
We report the quantitative results of our denoising experiment
in Table~\ref{tab:denoising_results}.
All the methods we compared were able to significantly denoise the input image,
improving the noisy input by at least 6.5 dB.
The reported SNRs are lower than would be typical
for natural image with this level of noise
(c.f. \cite{zhang_gaussian_2017}, where results are around 30~dB);
this is both because we report SNR rather than PSNR
and because our dataset lacks the large,
smooth areas typically found in high resolution natural images.
Parameter sweeps for BM3D and the TV-, and DCT-based methods took under ten minutes;
training for the unsupervised learning method took minutes;
and training for the supervised method took eight hours.
For all methods, reconstructions took less than one minute.

The proposed supervised learning-based method gave the best result,
followed by unsupervised learning, BM3D, TV, and DCT-based denoising.
The strong performance of the proposed  method
shows that learning was successful and that the learned analysis operator generalizes to unseen images.
That the supervised method outperforms TV is notable,
because, naively, 
we might have guessed that anisotropic TV is an optimal sparsifying transform for these images
(because they comprise only horizontal and vertical edges).
Because BM3D and the proposed method exploit different types of image structure (self-similarity vs local sparsity),
we conjecture that combining the methods by performing supervised analysis operator learning on matched or grouped patches could further improve performance.
Reference \cite{wen_vidosat_2019} shows that such an approach works in the unsupervised case.

\setuldepth{TV}
\begin{table}[!htpb]
    \centering
    \caption{Results of the denoising experiment.
    See Section~\ref{sec:meth_compared} for method details.}
    \begin{tabular}{@{}r r lllll@{}}
    \toprule
         & \ul{input}  & \ul{BM3D} &   \ul{TV} & \ul{DCT}  & \ul{unsupervised} & \ul{supervised (proposed)}  \\
     \textbf{testing SNR (dB)}  & 15.29  & 22.82 &22.59 & 21.90  & 22.84 & \textbf{23.16} \\
     \bottomrule
    \end{tabular}
    \label{tab:denoising_results}
\end{table}

\setlength{\imwidth}{0.16\linewidth}
\begin{figure}[!htbp]
    \captionsetup[subfigure]{justification=centering}
    \captionsetup[subfigure]{aboveskip=0.5pt,belowskip=-4pt}
    \centering
    \begin{subfigure}{\imwidth}
        \includegraphics[width=\textwidth]{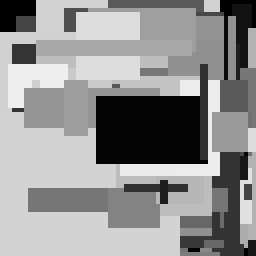}
        \caption{truth, $\infty$ dB}
        \label{fig:denoising_results_gt}
    \end{subfigure}\hfill
    \begin{subfigure}{\imwidth} 
        \includegraphics[width=\textwidth]{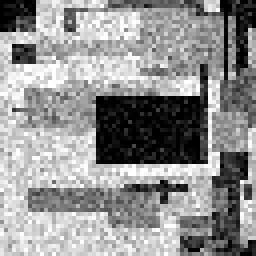}
        \caption{noisy, 16.2~dB}
        \label{fig:denoising_results_noisy}        
    \end{subfigure}\hfill
    \begin{subfigure}{2\imwidth}
        \includegraphics[width=.5\textwidth]{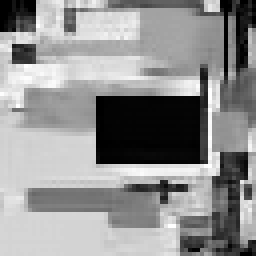}%
        \includegraphics[width=.5\textwidth]{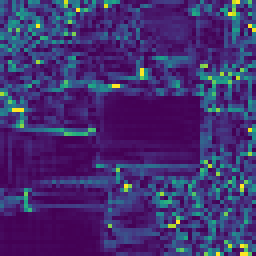}
        \caption{BM3D, 22.8~dB}
    \end{subfigure}\hfill
    \begin{subfigure}{2\imwidth}
        \includegraphics[width=.5\textwidth]{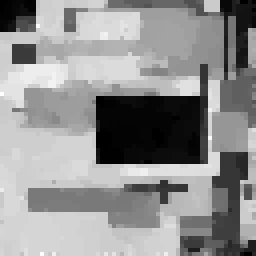}%
        \includegraphics[width=.5\textwidth]{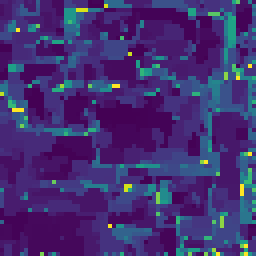}
        \caption{TV 23.2~dB}
    \end{subfigure}\vspace{.25cm}\\
    \begin{subfigure}{2\imwidth}
        \includegraphics[width=.5\textwidth]{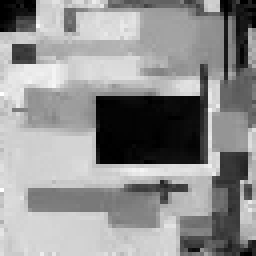}%
        \includegraphics[width=.5\textwidth]{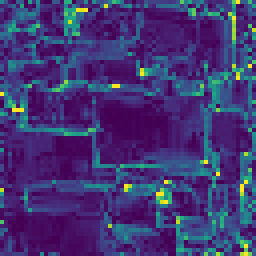}
        \caption{DCT, 22.1~dB}
    \end{subfigure}\hfill
    \begin{subfigure}{2\imwidth}
        \includegraphics[width=.5\textwidth]{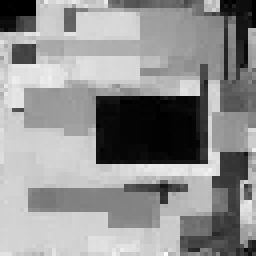}%
        \includegraphics[width=.5\textwidth]{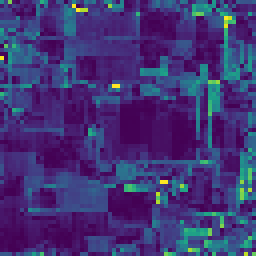}
        \caption{unsupervised, 22.9~dB}
    \end{subfigure}\hfill
    \begin{subfigure}{2\imwidth}
        \includegraphics[width=.5\textwidth]{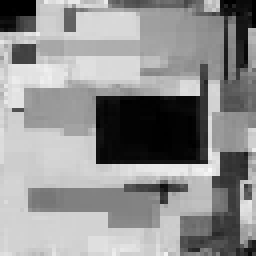}%
        \includegraphics[width=.5\textwidth]{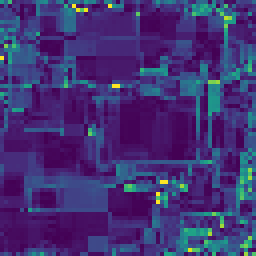}
        \caption{supervised, 23.2~dB}
    \end{subfigure}%
        \vspace{0.25cm}\\
    \hspace*{.5cm}%
    \begin{subfigure}{2.6\imwidth}%
        \begin{tikzpicture}%
            \draw (0, 0) node[inner sep=0] {\frame{\includegraphics[width=2.2\imwidth]{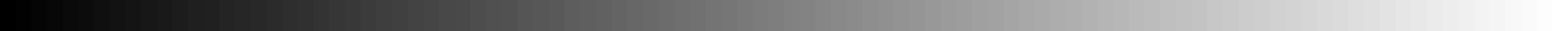}}};
            \draw (-2.5, -.25) node {0};
            \draw (2.5, -.25) node {1};
        \end{tikzpicture}%
    \end{subfigure}%
    \hspace*{\fill}%
    \begin{subfigure}{2.6\imwidth}%
        \begin{tikzpicture}%
            \draw (0, 0) node[inner sep=0] {\frame{\includegraphics[width=2.2\imwidth]{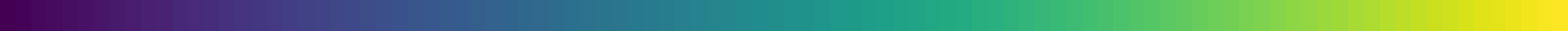}}};
            \draw (-2.5, -.25) node {0};
            \draw (2.5, -.25) node {0.2};
        \end{tikzpicture}%
    \end{subfigure}%
    \hspace*{.5cm}%
    \caption{Denoising results on a testing image.
    Each result is shown with the corresponding absolute error map
    and each subcaption reports the SNR on this image.}
    \label{fig:denoising_results}
\end{figure}
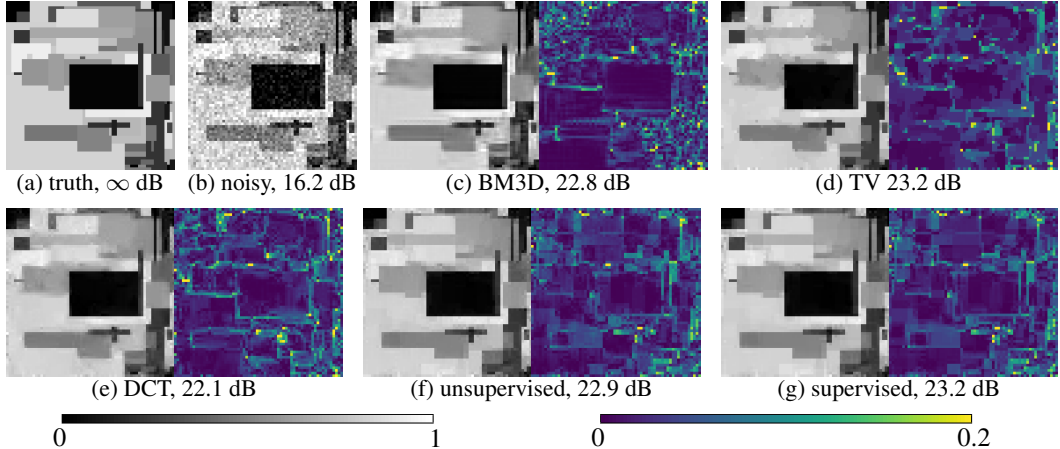

\setlength{\imwidth}{0.16\linewidth} 
\begin{figure}[!htbp]
    \centering
    \captionsetup[subfigure]{justification=centering}
    \captionsetup[subfigure]{aboveskip=0.5pt,belowskip=-4pt}    
    %
    %
    \begin{subfigure}{2\imwidth} 
    \includegraphics[width=\imwidth]{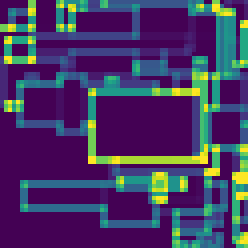}\hfill 
    \begin{subfigure}[b]{\imwidth}%
        \hspace*{\fill}%
        \hspace{.3\textwidth}\hfill%
        \includegraphics[width=.3\textwidth]{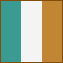}\hfill%
        \includegraphics[width=.3\textwidth]{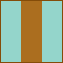}%
        \hspace*{\fill}\vspace{.08cm}\\\hspace*{\fill}%
        \includegraphics[width=.3\textwidth]{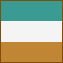}\hfill%
        \includegraphics[width=.3\textwidth]{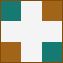}\hfill%
        \includegraphics[width=.3\textwidth]{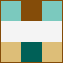}%
        \hspace*{\fill}\vspace{.08cm}\\\hspace*{\fill}%
        \includegraphics[width=.3\textwidth]{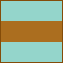}\hfill%
        \includegraphics[width=.3\textwidth]{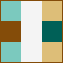}\hfill%
        \includegraphics[width=.3\textwidth]{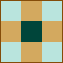}%
        \hspace*{\fill}%
    \end{subfigure}
    \caption{DCT}
    \label{fig:response_DCT}%
    \end{subfigure}%
    \hspace*{\fill}%
    %
    %
    \begin{subfigure}{2\imwidth} 
        \includegraphics[width=\imwidth]{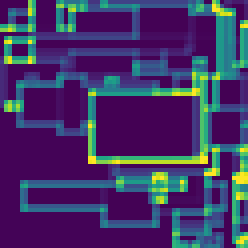}\hfill
        \begin{subfigure}[b]{.5\textwidth}%
        \hspace*{\fill}%
        \hspace{.3\textwidth}\hfill%
        \includegraphics[width=.3\textwidth]{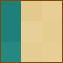}\hfill%
        \includegraphics[width=.3\textwidth]{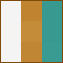}%
        \hspace*{\fill}\vspace{.08cm}\\\hspace*{\fill}%
        \includegraphics[width=.3\textwidth]{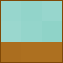}\hfill%
        \includegraphics[width=.3\textwidth]{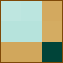}\hfill%
        \includegraphics[width=.3\textwidth]{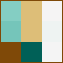}%
        \hspace*{\fill}\vspace{.08cm}\\\hspace*{\fill}%
        \includegraphics[width=.3\textwidth]{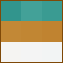}\hfill%
        \includegraphics[width=.3\textwidth]{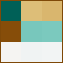}\hfill%
        \includegraphics[width=.3\textwidth]{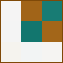}%
        \hspace*{\fill}
        \label{fig:response_unsupervised_filters}%
    \end{subfigure}%
        \caption{unsupervised}%
        \label{fig:response_unsupervised}%
    \end{subfigure}%
    \hspace*{\fill}%
    %
    %
    \begin{subfigure}{2\imwidth} 
        \includegraphics[width=\imwidth]{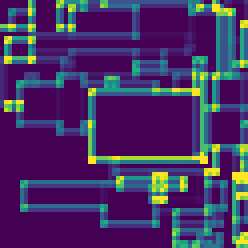}\hfill%
         \begin{subfigure}[b]{.5\textwidth}%
        \hspace*{\fill}%
        \hspace{.3\textwidth}\hfill%
        \includegraphics[width=.3\textwidth]{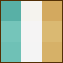}\hfill%
        \includegraphics[width=.3\textwidth]{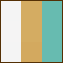}%
        \hspace*{\fill}\vspace{.08cm}\\\hspace*{\fill}%
        \includegraphics[width=.3\textwidth]{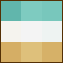}\hfill%
        \includegraphics[width=.3\textwidth]{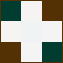}\hfill%
        \includegraphics[width=.3\textwidth]{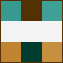}%
        \hspace*{\fill}\vspace{.08cm}\\\hspace*{\fill}%
        \includegraphics[width=.3\textwidth]{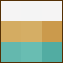}\hfill%
        \includegraphics[width=.3\textwidth]{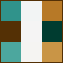}\hfill%
        \includegraphics[width=.3\textwidth]{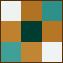}%
        \hspace*{\fill}%
        \end{subfigure}%
        \caption{supervised}
        \label{fig:response_supervised}%
    \end{subfigure}%
    \vspace{0.25cm}\\
        \hspace*{.5cm}%
    \begin{subfigure}{2.6\imwidth}%
        \begin{tikzpicture}%
            \draw (0, 0) node[inner sep=0] {\frame{\includegraphics[width=2.2\imwidth]{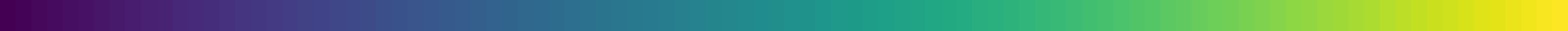}}};
            \draw (-2.5, -.25) node {0};
            \draw (2.5, -.25) node {2};
        \end{tikzpicture}%
    \end{subfigure}%
\hspace*{\fill}%
    \begin{subfigure}{2.6\imwidth}%
        \begin{tikzpicture}%
            \draw (0, 0) node[inner sep=0] {\frame{\includegraphics[width=2.2\imwidth]{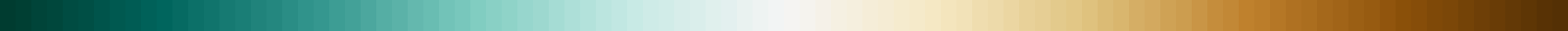}}};
            \draw (-2.5, -.25) node {-0.7};
            \draw (0, -.25) node {0};
            \draw (2.5, -.25) node {0.7};
        \end{tikzpicture}%
    \end{subfigure}%
    \hspace*{.7cm}%
    \caption{Summed absolute filter responses (left) and filters (right).}
    \label{fig:response}
\end{figure}

We can also explore the results by looking at the output from
each algorithm on a single image (Figure~\ref{fig:denoising_results}).
Qualitatively,
the BM3D and DCT results look oversmoothed,
probably as a result of patch averaging and
penalizing high frequencies, respectively.
The TV result has the characteristic 
stairstep pattern.
The results for both the unsupervised and supervised methods
are sharper than the DCT and BM3D results and avoid the stairstep pattern of TV,
but they otherwise look similar.
The error map for the supervised method looks less noisy
than for the unsupervised method,
which suggests that the supervised method does a better job
of smoothing flat areas in the image.

As another way to investigate our results,
we display the filters 
and the corresponding aggregate filter response map 
(sum of the absolute filter responses over channels) 
on a clean testing image 
for the DCT-, unsupervised learning-,
and supervised learning-based methods
(Figure~\ref{fig:response}).
Both the unsupervised and supervised methods are initialized with DCT filters,
and both make visible changes to them during training.
Both approaches also result in highly-structured filters.
However,
the unsupervised learning filters are (by construction) orthonormal,
while the supervised ones are not.
The supervised method seems to use this flexibility in two ways.
First, 
the filters are weighted differently,
with norms ranging from 0.75 to 1.54.
This effect is apparent in Figure~\ref{fig:response_supervised},
where the filters in the first row and first column  
are less saturated than the others.
Second,
the filters are not  orthogonal
(e.g., note the pair of filters in the first row in Figure~\ref{fig:response_supervised}).
From looking at the aggregate response maps,
we see that the effect of these differences is that aggregate filter responses
for the supervised method are are smaller on edges 
and larger on corners
than for either of the other methods.

\section{Conclusions and Future Work}
We have presented a new approach for learning a sparsifying analysis operator
for use in variational denoising.
The main contribution of the work is an expression for the gradient of the training objective $\eqref{eq:main}$
that does not rely on relaxation to a differentiable sparsity penalty.

We are interested in adapting this approach 
in two main directions.
First, we want to include a linear forward model in the objective function in \eqref{eq:main_lower},
making the approach applicable to the general class of linear inverse problems.
The main challenge in this generalization is that
the invertibility of the KKT matrix would depend on the properties forward operator.
Second, we would like to extend the approach to the case where $\W$ is a nonlinear function, e.g., a CNN.
We suspect that for a ReLU CNN,
our the KKT approach can still be used,
with the change that the sign pattern must be specified at each ReLU activation function.
Moving to a CNN-based regularizer also complicates
solving the regularized reconstruction problem itself,
because the proximal map used inside ADMM would no longer exist.
On the experimental side,
we would like to validate the approach on natural images
and make a thorough comparison to recent, CNN-based approaches for denoising,
e.g. \cite{zhang_gaussian_2017}.
These experiments may provide insight into how much of the performance of these methods come from their architecture
and how much comes from the supervised learning per se.


\section*{Broader Impact}
While our work is primarily of a theoretical and exploratory nature (using no real data),
it certainly may eventually lead to societal impacts.
We expect these to be mainly positive:
our work aims to improve the quality of image reconstructions, 
which, e.g., could help make MRI images clearer,
allowing doctors to make better clinical decisions.
However, this technology could also be used in negative ways,
e.g., to create better images from intrusive surveillance systems.
Using learning in these systems has its own pitfalls.
Returning to the MRI example, training data from minority populations may be scarce, resulting in systems that produce spurious results for minority patients at a higher rate than for patients in the majority.
As compared to highly-parameterized systems like those used in deep learning,
our shallow learning approach requires much less training data to perform well,
which may help ameliorate this type of disparity.


\bibliography{refs_mike}

\begin{thebibliography}{39}
\providecommand{\natexlab}[1]{#1}
\def\url#1{}
\csname url@samestyle\endcsname
\providecommand{\newblock}{\relax}
\providecommand{\bibinfo}[2]{#2}
\providecommand{\BIBentrySTDinterwordspacing}{\spaceskip=0pt\relax}
\providecommand{\BIBentryALTinterwordstretchfactor}{4}
\providecommand{\BIBentryALTinterwordspacing}{\spaceskip=\fontdimen2\font plus
\BIBentryALTinterwordstretchfactor\fontdimen3\font minus
  \fontdimen4\font\relax}
\providecommand{\BIBforeignlanguage}[2]{{%
\expandafter\ifx\csname l@#1\endcsname\relax
\typeout{** WARNING: IEEEtranN.bst: No hyphenation pattern has been}%
\typeout{** loaded for the language `#1'. Using the pattern for}%
\typeout{** the default language instead.}%
\else
\language=\csname l@#1\endcsname
\fi
#2}}
\providecommand{\BIBdecl}{\relax}
\BIBdecl

\bibitem[Bertero and Boccacci(1998)]{bertero_introduction_1998}
M.~Bertero and P.~Boccacci, \emph{Introduction to inverse problems in
  imaging}.\hskip 1em plus 0.5em minus 0.4em\relax Philadelphia, Pa: Institute
  of Physics Publishing, 1998.

\bibitem[McCann and Unser(2019)]{mccann_biomedical_2019}
M.~T. McCann and M.~Unser, ``Biomedical image reconstruction: From the
  foundations to deep neural networks,'' \emph{Foundations and Trends in Signal
  Processing}, vol.~13, no.~3, pp. 283--359, Dec. 2019.

\bibitem[Ravishankar et~al.(2020)Ravishankar, Ye, and
  Fessler]{ravishankar_image_2020}
S.~Ravishankar, J.~C. Ye, and J.~A. Fessler, ``Image reconstruction: From
  sparsity to data-adaptive methods and machine learning,'' \emph{Proceedings
  of the {IEEE}}, vol. 108, no.~1, pp. 86--109, Jan. 2020.

\bibitem[Tropp(2004)]{tropp_greed_2004}
J.~Tropp, ``Greed is good: Algorithmic results for sparse approximation,''
  \emph{{IEEE} Transactions on Information Theory}, vol.~50, no.~10, pp.
  2231--2242, Oct. 2004.

\bibitem[Rubinstein et~al.(2013)Rubinstein, Peleg, and
  Elad]{rubinstein_analysis_2013}
R.~Rubinstein, T.~Peleg, and M.~Elad, ``Analysis k-{SVD}: A dictionary-learning
  algorithm for the analysis sparse model,'' \emph{{IEEE} Transactions on
  Signal Processing}, vol.~61, no.~3, pp. 661--677, Feb. 2013.

\bibitem[Ravishankar and Bresler(2013)]{ravishankar_learning_2013}
S.~Ravishankar and Y.~Bresler, ``Learning sparsifying transforms,''
  \emph{{IEEE} Transactions on Signal Processing}, vol.~61, no.~5, pp.
  1072--1086, Mar. 2013.

\bibitem[Candes et~al.(2006)Candes, Romberg, and Tao]{candes_robust_2006}
E.~Candes, J.~Romberg, and T.~Tao, ``Robust uncertainty principles: exact
  signal reconstruction from highly incomplete frequency information,''
  \emph{{IEEE} Transactions on Information Theory}, vol.~52, no.~2, pp.
  489--509, Feb. 2006.

\bibitem[Tosic and Frossard(2011)]{tosic_dictionary_2011}
I.~Tosic and P.~Frossard, ``Dictionary learning,'' \emph{IEEE Signal Processing
  Magazine}, vol.~28, no.~2, pp. 27--38, Mar. 2011.

\bibitem[McCann et~al.(2017)McCann, Jin, and Unser]{mccann_convolutional_2017}
M.~T. McCann, K.~H. Jin, and M.~Unser, ``Convolutional neural networks for
  inverse problems in imaging: A review,'' \emph{IEEE Signal Processing
  Magazine}, vol.~34, no.~6, pp. 85--95, Nov. 2017.

\bibitem[Ongie et~al.(2020)Ongie, Jalal, Metzler, Baraniuk, Dimakis, and
  Willett]{ongie_deep_2020}
G.~Ongie, A.~Jalal, C.~A. Metzler, R.~G. Baraniuk, A.~G. Dimakis, and
  R.~Willett, ``Deep learning techniques for inverse problems in imaging,''
  \emph{arXiv:2005.06001 [eess.IV]}, May 2020.

\bibitem[Hel-Or and Shaked(2008)]{helor_discriminative_2008}
Y.~Hel-Or and D.~Shaked, ``A discriminative approach for wavelet denoising,''
  \emph{{IEEE} Transactions on Image Processing}, vol.~17, no.~4, pp. 443--457,
  Apr. 2008.

\bibitem[Shtok et~al.(2013)Shtok, Elad, and Zibulevsky]{shtok_learned_2013}
J.~Shtok, M.~Elad, and M.~Zibulevsky, ``Learned shrinkage approach for low-dose
  reconstruction in computed tomography,'' \emph{International Journal of
  Biomedical Imaging}, vol. 2013, pp. 1--20, Jun. 2013.

\bibitem[Kamilov and Mansour(2016)]{kamilov_learning_2016}
U.~S. Kamilov and H.~Mansour, ``Learning optimal nonlinearities for iterative
  thresholding algorithms,'' \emph{{IEEE} Signal Processing Letters}, vol.~23,
  no.~5, pp. 747--751, May 2016.

\bibitem[Nguyen et~al.(2018)Nguyen, Bostan, and Unser]{nguyen_learning_2018}
H.~Q. Nguyen, E.~Bostan, and M.~Unser, ``Learning convex regularizers for
  optimal bayesian denoising,'' \emph{{IEEE} Transactions on Signal
  Processing}, vol.~66, no.~4, pp. 1093--1105, Feb. 2018.

\bibitem[Al-Shabili et~al.(2020)Al-Shabili, Mansour, and
  Boufounos]{alshabili_learning_2020}
A.~H. Al-Shabili, H.~Mansour, and P.~T. Boufounos, ``Learning plug-and-play
  proximal quasi-newton denoisers,'' in \emph{2020 IEEE International
  Conference on Acoustics, Speech, and Signal Processing Proceedings},
  Barcelona, Spain, May 2020.

\bibitem[Gupta et~al.(2018)Gupta, Jin, Nguyen, McCann, and
  Unser]{gupta_cnn_2018}
H.~Gupta, K.~H. Jin, H.~Q. Nguyen, M.~T. McCann, and M.~Unser, ``{CNN}-based
  projected gradient descent for consistent {CT} image reconstruction,''
  \emph{IEEE Transactions on Medical Imaging}, vol.~37, no.~6, pp. 1440--1453,
  Jun. 2018.

\bibitem[Aggarwal et~al.(2019)Aggarwal, Mani, and Jacob]{aggarwal_modl_2019}
H.~K. Aggarwal, M.~P. Mani, and M.~Jacob, ``{MoDL}: Model-based deep learning
  architecture for inverse problems,'' \emph{{IEEE} Transactions on Medical
  Imaging}, vol.~38, no.~2, pp. 394--405, Feb. 2019.

\bibitem[Gribonval and Machart(2013)]{gribonval_reconciling_2013}
\BIBentryALTinterwordspacing
R.~Gribonval and P.~Machart, ``Reconciling "priors" \& "priors" without
  prejudice?'' in \emph{Advances in Neural Information Processing Systems 26},
  Dec. 2013, vol.~2, pp. 2193--2201.
  \url{http://papers.nips.cc/paper/4868-reconciling-priors-priors-without-prejudice.pdf}
\BIBentrySTDinterwordspacing

\bibitem[Peyr{\'e} and Fadili(2011)]{peyre_learning_2011}
\BIBentryALTinterwordspacing
G.~Peyr{\'e} and J.~M. Fadili, ``Learning analysis sparsity priors,'' in
  \emph{Sampling Theory and Applications}, Singapore, Singapore, May 2011,
  p.~4.  \url{https://hal.archives-ouvertes.fr/hal-00542016}
\BIBentrySTDinterwordspacing

\bibitem[Mairal et~al.(2012)Mairal, Bach, and Ponce]{mairal_task_2012}
J.~Mairal, F.~Bach, and J.~Ponce, ``Task-driven dictionary learning,''
  \emph{{IEEE} Transactions on Pattern Analysis and Machine Intelligence},
  vol.~34, no.~4, pp. 791--804, Apr. 2012.

\bibitem[Sprechmann et~al.(2013)Sprechmann, Litman, Ben~Yakar, Bronstein, and
  Sapiro]{sprechmann_supervised_2013}
\BIBentryALTinterwordspacing
P.~Sprechmann, R.~Litman, T.~Ben~Yakar, A.~M. Bronstein, and G.~Sapiro,
  ``Supervised sparse analysis and synthesis operators,'' in \emph{Advances in
  Neural Information Processing Systems 26}, 2013, pp. 908--916.
  \url{http://papers.nips.cc/paper/5002-supervised-sparse-analysis-and-synthesis-operators.pdf}
\BIBentrySTDinterwordspacing

\bibitem[Chen et~al.(2014{\natexlab{a}})Chen, Pock, and
  Bischof]{chen_learning_2014}
Y.~Chen, T.~Pock, and H.~Bischof, ``Learning $\ell_1$-based analysis and
  synthesis sparsity priors using bi-level optimization,''
  \emph{arXiv:1401.4105 [cs.CV]}, Jan. 2014.

\bibitem[Chen et~al.(2014{\natexlab{b}})Chen, Ranftl, and
  Pock]{chen_insights_2014}
Y.~Chen, R.~Ranftl, and T.~Pock, ``Insights into analysis operator learning:
  From patch-based sparse models to higher order {MRFs},'' \emph{{IEEE}
  Transactions on Image Processing}, vol.~23, no.~3, pp. 1060--1072, Mar. 2014.

\bibitem[Boyd and Vandenberghe(2004)]{boyd_convex_2004}
S.~Boyd and L.~Vandenberghe, \emph{Convex Optimization}.\hskip 1em plus 0.5em
  minus 0.4em\relax Cambridge University Press, 2004.

\bibitem[Golub and Pereyra(1973)]{golub_differentiation_1973}
\BIBentryALTinterwordspacing
G.~H. Golub and V.~Pereyra, ``The differentiation of pseudo-inverses and
  nonlinear least squares problems whose variables separate,'' \emph{SIAM
  Journal on Numerical Analysis}, vol.~10, no.~2, pp. 413--432, 1973.
  \url{http://www.jstor.org/stable/2156365}
\BIBentrySTDinterwordspacing

\bibitem[Paszke et~al.(2019)Paszke, Gross, Massa, Lerer, Bradbury, Chanan,
  Killeen, Lin, Gimelshein, Antiga, Desmaison, Kopf, Yang, DeVito, Raison,
  Tejani, Chilamkurthy, Steiner, Fang, Bai, and Chintala]{paszke_pytorch_2019}
\BIBentryALTinterwordspacing
A.~Paszke, S.~Gross, F.~Massa, A.~Lerer, J.~Bradbury, G.~Chanan, T.~Killeen,
  Z.~Lin, N.~Gimelshein, L.~Antiga, A.~Desmaison, A.~Kopf, E.~Yang, Z.~DeVito,
  M.~Raison, A.~Tejani, S.~Chilamkurthy, B.~Steiner, L.~Fang, J.~Bai, and
  S.~Chintala, ``Pytorch: An imperative style, high-performance deep learning
  library,'' in \emph{Advances in Neural Information Processing Systems 32},
  2019, pp. 8024--8035.
  \url{http://papers.neurips.cc/paper/9015-pytorch-an-imperative-style-high-performance-deep-learning-library.pdf}
\BIBentrySTDinterwordspacing

\bibitem[Minka(2000)]{minka_old_2000}
\BIBentryALTinterwordspacing
T.~P. Minka, ``Old and new matrix algebra useful for statistics,'' MIT Media
  Lab, Tech. Rep., 2000.  \url{https://tminka.github.io/papers/matrix/}
\BIBentrySTDinterwordspacing

\bibitem[Boyd et~al.(2011)Boyd, Parikh, Chu, Peleato, and
  Eckstein]{boyd_distributed_2011}
S.~Boyd, N.~Parikh, E.~Chu, B.~Peleato, and J.~Eckstein, ``Distributed
  optimization and statistical learning via the alternating direction method of
  multipliers,'' \emph{Foundations and Trends in Machine Learning}, vol.~3,
  no.~1, pp. 1--122, Jan. 2011.

\bibitem[Shewchuk(1994)]{shewchuk_introduction_1994}
\BIBentryALTinterwordspacing
J.~Shewchuk, ``An introduction to the conjugate gradient method without the
  agonizing pain,'' Carnegie Mellon University, Tech. Rep., 1994.
  \url{http://portal.acm.org/citation.cfm?id=865018}
\BIBentrySTDinterwordspacing

\bibitem[Kingma and Ba(2014)]{kingma_adam_2014}
D.~P. Kingma and J.~Ba, ``Adam: A method for stochastic optimization,''
  \emph{arXiv:1412.6980 [cs.LG]}, Dec. 2014.

\bibitem[Avriel(2003)]{avriel_nonlinear_2003}
M.~Avriel, \emph{Nonlinear Programming: Analysis and Methods}.\hskip 1em plus
  0.5em minus 0.4em\relax Mineola, NY: Dover Publications, Jun. 2003.

\bibitem[Lee et~al.(2001)Lee, Mumford, and Huang]{lee_occlusion_2001}
A.~Lee, D.~Mumford, and J.~Huang, ``Occlusion models for natural images: a
  statistical study of a scale invariant dead leaves model,''
  \emph{International Journal of Computer Vision}, vol.~41, pp. 35--59, Jan.
  2001.

\bibitem[Makinen et~al.(2019)Makinen, Azzari, and Foi]{makinen_exact_2019}
Y.~Makinen, L.~Azzari, and A.~Foi, ``Exact transform-domain noise variance for
  collaborative filtering of stationary correlated noise,'' in \emph{2019
  {IEEE} International Conference on Image Processing Proceedings}, Taipei,
  Taiwan, Sep. 2019.

\bibitem[Yaghoobi et~al.(2012)Yaghoobi, Nam, Gribonval, and
  Davies]{yaghoobi_noise_2012}
M.~Yaghoobi, S.~Nam, R.~Gribonval, and M.~E. Davies, ``Noise aware analysis
  operator learning for approximately cosparse signals,'' in \emph{2012 {IEEE}
  International Conference on Acoustics, Speech and Signal Processing
  ({ICASSP})}.\hskip 1em plus 0.5em minus 0.4em\relax {IEEE}, Mar. 2012.

\bibitem[Ravishankar and Bresler(2015)]{ravishankar_sparsifying_2015}
S.~Ravishankar and Y.~Bresler, ``Sparsifying transform learning with efficient
  optimal updates and convergence guarantees,'' \emph{{IEEE} Transactions on
  Signal Processing}, vol.~63, no.~9, pp. 2389--2404, May 2015.

\bibitem[Smith et~al.(2017)Smith, Kindermans, Ying, and Le]{smith_dont_2017}
S.~L. Smith, P.-J. Kindermans, C.~Ying, and Q.~V. Le, ``Don't decay the
  learning rate, increase the batch size,'' \emph{arXiv:1711.00489 [cs.LG]},
  Nov. 2017.

\bibitem[Pineau and Sinha(2020)]{sinha_designing_2020}
\BIBentryALTinterwordspacing
J.~Pineau and K.~Sinha, ``Designing the reproducibility program for {NeurIPS}
  2020,'' medium.com, Apr. 2020.
  \url{https://medium.com/@NeurIPSConf/designing-the-reproducibility-program-for-neurips-2020-7fcccaa5c6ad}
\BIBentrySTDinterwordspacing

\bibitem[Zhang et~al.(2017)Zhang, Zuo, Chen, Meng, and
  Zhang]{zhang_gaussian_2017}
K.~Zhang, W.~Zuo, Y.~Chen, D.~Meng, and L.~Zhang, ``Beyond a gaussian denoiser:
  Residual learning of deep {CNN} for image denoising,'' \emph{{IEEE}
  Transactions on Image Processing}, vol.~26, no.~7, pp. 3142--3155, Jul. 2017.

\bibitem[Wen et~al.(2019)Wen, Ravishankar, and Bresler]{wen_vidosat_2019}
B.~Wen, S.~Ravishankar, and Y.~Bresler, ``{VIDOSAT}: High-dimensional
  sparsifying transform learning for online video denoising,'' \emph{{IEEE}
  Transactions on Image Processing}, vol.~28, no.~4, pp. 1691--1704, Apr. 2019.

\end{thebibliography}

\end{document}